\documentclass[aps,prl,twocolumn,groupedaddress]{revtex4}
\usepackage{graphicx}
\usepackage{amsmath}
\usepackage{amssymb}
\usepackage{amsfonts}
\usepackage{epsfig}
\begin{document}

\newcommand{\lindisp}{\beta}
\newcommand{\sbno}{n}

\title{Observation of Nonlocal Modulation with Entangled Photons}


\author{S. Sensarn}
\email{sensarn@stanford.edu}
\author{G.Y. Yin}
\author{S.E. Harris}
\affiliation{Edward L. Ginzton Laboratory, Stanford University, Stanford, California 94305}


\date{\today}

\begin{abstract}
We demonstrate a new type of quantum mechanical correlation where phase modulators at distant locations, acting on the photons of an entangled pair, interfere to determine the apparent depth of modulation.  When the modulators have the same phase, the modulation depth doubles; when oppositely phased, the modulators negate each other.
\end{abstract}

\pacs{}

\maketitle 

A nonlinear crystal pumped by a monochromatic laser may generate time-energy entangled photon pairs through the process of spontaneous parametric down conversion \cite{Shih03}.  These pairs may be spatially separated into two channels and their frequencies and arrival times at distant detectors measured.  If Observer A measures the frequency of her photon, she knows, by energy conservation, the exact frequency of the photon measured by Observer B.  If, instead, she measures the arrival time of her photon, she can predict the detection time measured by Observer B to within a small window that varies inversely with the photon bandwidth \cite{Strekalov05} and may be lengthened by dispersion \cite{Valencia02}.  The ability to measure relative time and relative frequency, with accuracies not limited by time-energy uncertainty, is the hallmark of time-energy entanglement \cite{Mancini02, Howell04, IAK06}.  

An important consequence of time-energy entanglement, first noted by Franson \cite{FransonNLD, Franson02} and observed by  Brendel et al. \cite{Brendel98}, is nonlocal cancellation of dispersion. When the photons of an entangled pair are sent through different channels having arbitrary dispersions, the dispersion in one channel may be negated by dispersion of the opposite sign in the other channel. This effect results from quantum interference and has no classical analog.  As always, nonlocal effects do not imply transmission of classical information.

\begin{figure}
\includegraphics [scale=0.38, clip]{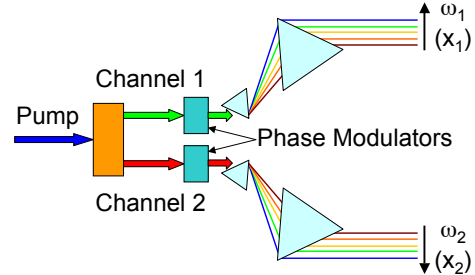}
\caption{(color online).  Schematic of nonlocal modulation.}
\label{setup_basic}
\end{figure}

In this Letter, we report the first observation of a time-frequency analog to nonlocal dispersion cancellation and term this effect as nonlocal modulation \cite{HarrisNLM}. Spontaneous down conversion is used to generate entangled photons that each have a spectral width of about 280~$\text{cm}^{-1}$. Sinusoidal phase modulators, each operating at a frequency of 30~GHz (1~$\text{cm}^{-1}$), are placed in the two beams, labeled Channel 1 and Channel 2 (see Fig.~\ref{setup_basic}). When either beam is dispersed by a prism or grating and viewed in the frequency domain (i.e. as a function of position), this modulation is completely hidden by the much broader spectral width of the photon. To observe the modulation, we correlate in the frequency (spatial) domain. In the absence of modulation, a photon detected at a particular position in Channel 1 coincides with a photon of frequency $\omega_2 =\omega_p - \omega_1$ in Channel 2, and the correlation in the frequency (spatial) domain is therefore a delta function $\delta(\omega_1 + \omega_2 - \omega_p)$. When synchronously driven modulators are placed in the signal and idler channels, this correlation becomes a distribution of discrete sidebands spaced by the modulation frequency. What is strange and interesting is that these distant modulators now act cumulatively.  For example, if the two identical modulators have opposite phase, they negate each other and act as if neither modulator were present.  Conversely, if operated with the same phase, they produce the same correlation as does  a single modulator with twice the modulation depth acting on only one of the photons.

To avoid confusion, we mention a different type of quantum interference discovered by Steinberg and colleagues \cite{Steinberg92, Steinberg92_2} that bears on Hong-Ou-Mandel interferometry. Here, because it is not possible to determine which photon passed through a dispersive medium, there is an interference of  Feynman paths, and  even-order dispersive terms are not seen by the interferometer. Recognizing the importance of time-frequency duality, Tsang and Psaltis have suggested the equivalent of the Steinberg interference in the time domain \cite{Psaltis05, Psaltis06}.  Since the photons meet on a single beam splitter in these examples, the interference is a local effect, and classical analogies have been demonstrated \cite{Kaltenbaek08, Kaltenbaek09}.

\begin{figure}
\includegraphics [scale=0.48, clip]{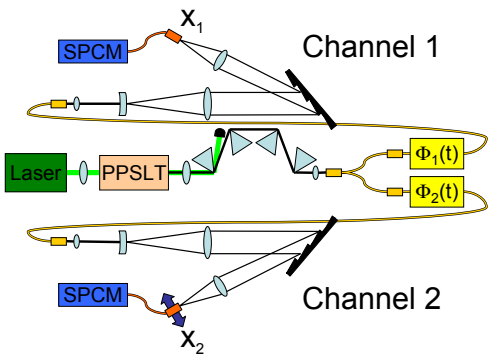}
\caption{(color online).  Experimental setup. $\Phi_1(t)$ and $\Phi_2(t)$ are 30 GHz sinusoidal modulators synchronously driven with variable relative phase (see text for details).}
\label{setup}
\end{figure}

A schematic of the experiment is shown in Fig.~\ref{setup}.  We pump a 20 mm long, periodically-poled, magnesium oxide-doped stoichiometric lithium tantalate crystal (PPSLT, HC Photonics Corp.) with 0.8~W from a 532 nm cw laser (Coherent Verdi V10).  The nonlinear crystal is phase matched to produce 32~nm bandwidth, degenerate photon pairs at 1064~nm.  All fields are polarized along the extraordinary axis of the crystal.  The generated photons are filtered from the strong 532~nm pumping beam using a four-prism setup and are then coupled into a polarization-maintaining fused-fiber beam splitter which diverts the photons into Channels 1 and 2 with equal probability.  The photons pass through identical sinusoidal phase modulators (EOSPACE) driven at 30~GHz with modulation depths of about 1.5 radians.  The relative phase between the modulators is controlled using a calibrated phase trimmer.  Following the modulators are identical monochromators, each having a linear dispersion of 210~GHz/mm and a Gaussian instrument response function with a FWHM bandwidth of 8.5~GHz.  To obtain frequency-domain correlations, we fix the output slit in Channel 1 at $x_1$ and scan the position $x_2$ in Channel 2.  The photons transmitted through the monochromator slits are coupled into multimode fibers and detected with single photon counting modules (SPCMs, id Quantique id400 and PerkinElmer SPCM-AQR-16-FC).

\begin{figure}[h]
\includegraphics [scale=0.63, clip]{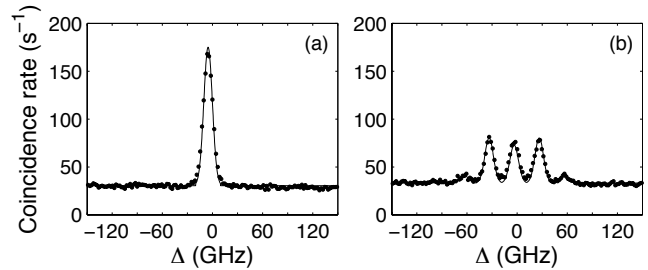}
\caption{Frequency correlation measurements (a) with both modulators turned off and (b) with the modulator in Channel 1 running at a modulation depth of 1.5.  Dots are data; curves are theoretical fits (see text).  All data is approximately shot-noise-limited. }
\label{data1}
\end{figure}

\begin{figure}[h]
\includegraphics [scale=0.63, clip]{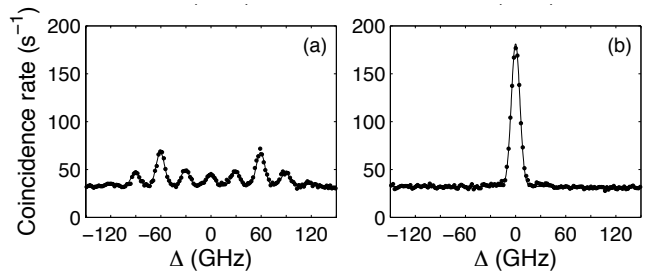}
\caption{As in Fig.~\ref{data1} but both modulators running (a) with the same phase and (b) with opposite phase. }
\label{data2}
\end{figure}

The primary experimental results of this work are shown in Figs.~\ref{data1} and \ref{data2}.  For each case, we set the monochromator slit in Channel 1 at an arbitrary position $x_1$ near the center of the generated 32~nm spectrum and leave the position of this slit fixed thereafter. The slit in Channel 2 is scanned over positions $x_2$, and the coincidence rate of the two detectors (with gate width $T = 1.25$~ns) is recorded as a function of this position.  For each position, the rate is averaged for 20~s.

With the pump frequency defined as $\omega_p$, and the position $x_2$ proportional to the frequency $\omega_2$, we express the coincidence rate as a function of relative frequency $\Delta \equiv \omega_2 - (\omega_p - \omega_1)$. The scale of the frequency axis is calibrated by measuring the sideband spacing of a single-mode 1064~nm laser modulated at 30~GHz, with the zero position chosen (at the start of the experiment) as the location of the correlation peak for unmodulated photon pairs.

Figure~\ref{data1} shows the experimental results without modulation and with modulation in a single channel.  In Fig.~\ref{data1}(a), both modulators are turned off by disconnecting their 30~GHz drive signals. As expected by energy conservation, a single correlation peak is observed.  In Figure~\ref{data1}(b), Channel 1 is phase modulated as $\exp[i \delta \cos(\omega_m t)]$ with a modulation depth of $\delta=1.5$, and Channel 2 is not modulated. The frequency correlation is now distributed over a set of sidebands, having Bessel function amplitudes $J_{n}^{2}(\delta)$, whose total area is equal to that of Fig.~\ref{data1}(a).

In Fig.~\ref{data2}(a), both modulators are turned on at a modulation depth of $\delta=1.5$, and the cable length is adjusted so that they have the same phase.  They now act cumulatively (constructively interfere) to produce a set of sidebands having a Bessel function distribution $J_{n}^{2}(2 \delta)$. The frequency-domain correlation function of two distant modulators is therefore the same as that which would be obtained by correlating an unmodulated photon with a photon modulated at twice the modulation depth. 

In Fig.~\ref{data2}(b), the modulators are run at the same depth as in the previous paragraph, but now the relative cable length is adjusted so that the modulators are run in phase opposition.  The modulators now destructively interfere, and no sidebands are visible. 
The solid curves in Figs.~\ref{data1} and \ref{data2} are theoretical fits to the data.  

The theory of nonlocal modulation as measured in the frequency domain has been developed by Harris \cite{HarrisNLM} for the case of frequency correlation using ideal detectors with perfect frequency (spatial) resolution. In the following paragraphs, we develop the theory to allow for finite-resolution monochromators and detectors having both arbitrary transmission functions and specified temporal gate widths.  Working in the Heisenberg picture, a nonlinear crystal of length $L$ is pumped by a monochromatic laser at frequency $\omega_p$.  A positive-frequency field operator $a(\omega, z)$, representing entangled photons, evolves inside the crystal and may be written in terms of an envelope $b(\omega, z)$ which varies slowly along the propagation direction: $a(\omega, z) = b(\omega, z) \exp[i k(\omega) z]$.  The propagation equations describing entangled photon generation are

\begin{eqnarray}
\!\!\!\!\!\! \frac{\partial b(\omega, z)}{\partial z} &=& i \kappa(\omega) b^\dag(\omega_p - \omega, z) \exp\left[ i \Delta k(\omega) z \right] \! , \nonumber \\
\!\!\!\!\!\! \frac{\partial b^\dag(\omega, z)}{\partial z} &=& -i \kappa^*(\omega) b(\omega_p - \omega, z) \exp\left[ -i \Delta k(\omega) z \right] \! .
\end{eqnarray}

\noindent
where $\kappa(\omega)$ and $\Delta k(\omega)$ are the coupling factor and wave-vector mismatch, respectively.  The solution for the output field at $z=L$, expressed in terms of the vacuum field $a_\text{vac}(\omega)$ at the input of the crystal, is

\begin{equation}
a_\text{out}(\omega) = A(\omega)a_\text{vac}(\omega) + B(\omega)a_\text{vac}^\dag(\omega_p - \omega) ,
\label{aout}
\end{equation}

\noindent
where, to preserve the commutation relations, the functions $A(\omega)$ and $B(\omega)$ satisfy $|A(\omega)|^2 - |B(\omega)|^2 = 1$ and $A(\omega) B(\omega_p - \omega) = B(\omega) A(\omega_p - \omega)$.

The time-domain output field operator is related to its frequency-domain counterpart [Eq.~(\ref{aout})] by the inverse Fourier transform, $a_\text{out}(t) = \int_{-\infty}^\infty{a_\text{out}(\omega) \exp(-i \omega t) d\omega}$, and is normalized so that the total rate of generated photons exiting the crystal is $R_\text{out} = \langle a_\text{out}^\dag(t) a_\text{out}(t) \rangle$.  The generated photons are separated into two channels, denoted as Channel~1 and Channel~2, using a 50/50 beam splitter.  The field operators at the outputs of the beam splitter are $a_{1}(t) = a_{2}(t)=\frac{1}{\sqrt{2}} a_\text{out}(t)$.  The photons are modulated by periodic phase modulators whose time-domain, Fourier-series transfer functions are $m_1(t) = \sum_k q_k \exp(-i k \omega_m t)$ in Channel~1 and  $m_2(t) = \sum_l r_l \exp(-i l \omega_m t)$ in Channel~2, with Fourier transforms $m_1(\omega) =  \sum_k q_k \delta(\omega - k \omega_m)$ and $m_2(\omega) = \sum_l r_l \delta(\omega - l \omega_m)$, respectively.  With the $*$ symbol denoting convolution, the frequency-domain modulated fields are $\tilde{a}_1(\omega) = a_1(\omega) * m_1(\omega)$ and $\tilde{a}_2(\omega) = a_2(\omega) * m_2(\omega)$.  Substituting $a_1(\omega)$, $a_2(\omega)$, $m_1(\omega)$, and $m_2(\omega)$ into the expressions for $\tilde{a}_1(\omega)$ and $\tilde{a}_2(\omega)$ yields

\begin{eqnarray}
\tilde{a}_1(\omega) &=& \frac{1}{\sqrt{2}} \sum_{k=-\infty}^{\infty} q_k \big[ A(\omega - k \omega_m) a_\text{vac}(\omega - k \omega_m) \nonumber \\
&& \null + B(\omega - k \omega_m) a_\text{vac}^\dag(\omega_p - \omega + k \omega_m) \big] , \nonumber \\
\tilde{a}_2(\omega) &=& \frac{1}{\sqrt{2}} \sum_{l=-\infty}^{\infty} r_l \big[ A(\omega - l \omega_m) a_\text{vac}(\omega - l \omega_m) \nonumber \\
&& \null + B(\omega - l \omega_m) a_\text{vac}^\dag(\omega_p - \omega + l \omega_m) \big] . \nonumber \\
\end{eqnarray}

The modulated photons are frequency correlated by passing each through identical monochromators whose output slits may be translated to select frequencies $\omega_1 = \lindisp x_1$ in Channel 1 and $\omega_2 = \lindisp x_2$ in Channel 2, where the constant $\lindisp$ is the linear dispersion of the grating systems.  The monochromators (spectral filters) have field transmission functions $H_1(\omega - \lindisp x_1)$ and $H_2(\omega -\lindisp x_2)$.  The filtered field operators in Channels 1 and 2 are $\tilde{a}_\text{1f}(\omega, x_1) = \tilde{a}_1(\omega) H_1(\omega - \lindisp x_1)$ and $\tilde{a}_\text{2f}(\omega, x_2) = \tilde{a}_2(\omega) H_2(\omega - \lindisp x_2)$, respectively.  The count rates at the outputs of the monochromators are given by $R_1(x_1) = \langle \tilde{a}_\text{1f}^\dag(t, x_1) \tilde{a}_\text{1f}(t, x_1) \rangle$ and $R_2(x_2) = \langle\tilde{ a}_\text{2f}^\dag(t, x_2) \tilde{a}_\text{2f}(t, x_2) \rangle$. These rates are

\begin{eqnarray}
R_1(x_1) \!\! &=& \!\! \frac{1}{4 \pi} \! \sum_{k = -\infty}^\infty \!\!\! |q_k|^2 \! \nonumber \\
&& \times \int_0^{\infty} \!\!\!\! \left| B(\omega - k \omega_m) \right| ^2 \left| H_1(\omega - \lindisp x_1) \right| ^2 \! d \omega , \nonumber \\
R_2(x_2) \!\! &=& \!\! \frac{1}{4 \pi} \! \sum_{l = -\infty}^\infty \!\!\! |r_l|^2 \! \nonumber \\
&& \times \int_0^{\infty} \!\!\!\! \left| B(\omega - l \omega_m) \right| ^2 \left| H_2(\omega - \lindisp x_2) \right| ^2 \! d \omega .
\label{RsRi}
\end{eqnarray}

Assuming a gate width $T$, the coincidence rate for the two detectors is related to the second-order Glauber correlation function $G^{(2)}(t_1, x_1, t_2, x_2) = \langle \tilde{a}_\text{2f}^\dag(t_2, x_2) \tilde{a}_\text{1f}^\dag(t_1, x_1) \tilde{a}_\text{1f}(t_1, x_1) \tilde{a}_\text{2f}(t_2, x_2) \rangle$.  With the assumption that the resolution of the monochromators is high, or equivalently that the filter widths are small (as compared to the modulation frequency $\omega_m$), it can be shown that the correlation function depends only on the difference of the arrival times $\tau = t_2 - t_1$, and the coincidence rate is

\begin{equation}
R_c(x_1, x_2) = \int_{-T/2}^{T/2} G^{(2)}(\tau, x_1, x_2) d\tau .
\label{Rc1}
\end{equation}

Equation~(\ref{Rc1}) may be expanded using Wick's theorem and shown to be given by

\begin{eqnarray}
\!\!\! R_c(x_1, x_2) \!\! &=& \!\! R_1(x_1) R_2(x_2) T  \nonumber \\
&& \!\! \null + \int_{-\infty}^\infty { \left| \sum_{k=-\infty}^{\infty}  \!\!\! q_k r_{\sbno - k} F_k(\tau, x_1, x_2) \right|^2 \!\!\!\! d\tau} \!,
\label{Rc2}
\end{eqnarray}

\noindent
where $\Delta = \lindisp(x_1 + x_2) - \omega_p$, $\sbno = \lfloor \Delta/\omega_m + \frac{1}{2} \rfloor$, and

\begin{eqnarray}
F_k(\tau, x_1, x_2) \!\! &=& \!\! \frac{1}{4 \pi} \int_0^\infty { \!\!\!\! A(\omega - k \omega_m) B(\omega_p - \omega + k \omega_m) } \nonumber \\
&& \!\! \null \times H_1(\omega - \lindisp x_1) H_2(\omega_p - \omega - \lindisp x_2 + \sbno \omega_m) \nonumber \\
&& \!\! \null \times \exp(i \omega \tau) d\omega .
\label{Fk}
\end{eqnarray}

The first term in Eq.~(\ref{Rc2}) is the result of accidental coincidences between unpaired photons in a gate width $T$. The second term is the coincidence rate between paired photons and captures the modulation effects described in this Letter. To obtain Eqs.~(\ref{RsRi})--(\ref{Fk}), we have assumed that the transmission widths of the monochromators are small as compared to the modulation frequency and large as compared to the inverse of the temporal gate width $T$. In our experiment these assumptions are satisfied by factors of 3.5 and 11, respectively. 

If we further assume that $A(\omega)$ and $B(\omega_p - \omega)$ are constant in the vicinity ($\pm$150~GHz in Figs.~\ref{data1} and \ref{data2}) of $\omega = \beta x_1$ and are equal to $A_0$ and $B_0$, respectively, then Eq.~(\ref{Rc2}) becomes

\begin{equation}
R_c(\Delta) = R_1 R_2 T + c_\sbno H^{(2)}(\sbno \omega_m - \Delta) , 
\label{Rc3}
\end{equation}

\noindent
where $H^{(2)}(\omega) = | H_1(\omega) |^2 * | H_2(\omega) |^2$, and

\addtocounter{equation}{-1}
\begin{subequations}
\begin{eqnarray}
R_1 &=& \frac{1}{4 \pi} |B_0|^2 \int_{-\infty}^{\infty} |H_1(\omega)|^2 d\omega , \label{R1} \\
R_2 &=& \frac{1}{4 \pi} |B_0|^2 \int_{-\infty}^{\infty} |H_2(\omega)|^2 d\omega , \label{R2} \\
c_\sbno &=& \frac{1}{8 \pi} \left| A_0 B_0 \!\! \sum_{k = -\infty}^{\infty} q_k r_{\sbno - k} \right|^2 \!\!\! . \label{cn}
\end{eqnarray}
\end{subequations}

The solid curves in Figs.~\ref{data1} and \ref{data2} are theoretical fits to the data using Eq.~(\ref{Rc3}) shifted horizontally so as to match center.  The Fourier series coefficients for sinusoidal phase modulators are Bessel functions with $q_k = J_k(-\delta_1)$ and $r_l = J_l(-\delta_2)$, where $\delta_1$ and $\delta_2$ are the modulation depths in Channels 1 and 2, respectively ($|\delta_1| = |\delta_2| = 1.5$ in our experiment).  We model the monochromator response functions in Channels 1 and 2 as Gaussians with FWHM bandwidths $\Gamma$: $H_1(\omega) = \alpha_1 \exp \left [-2 \ln(2) \omega^2/\Gamma^2 \right]$ and $H_2(\omega) = \alpha_2\exp \left [-2 \ln(2) \omega^2/\Gamma^2 \right ]$.  (The monochromator in Channel 1 is the mirror image of the one in Channel 2 which has a measured FWHM bandwidth of 8.5~GHz.)  The transfer functions include fitting parameters $\alpha_1$ and $\alpha_2$ used in Figs.~\ref{data1} and \ref{data2} to account for transmission losses and the difference in detection efficiencies of the photon counters.  

To obtain the constants $A_0$ and $B_0$, for each case in Figs.~\ref{data1} and \ref{data2}, we measure the average value of $R_2$ and use Eq.~(\ref{R2}) to calculate $|B_0|$. We obtain $|A_0|$ from the commutator-preserving condition $|A_0|^2 - |B_0|^2 = 1$.  For all curves, the fitting parameters are taken as $\alpha_1^2 = 1.20 \times 10^{-2}$ and $\alpha_2^2 = 5.59 \times 10^{-4}$. These values are in good agreement with loss measurements and estimates of the photon counter detection efficiency, where we note that the id400 detector in Channel 1 has a detection efficiency an order of magnitude larger than the SPCM-AQR-16-FC detector in Channel 2.

In summary, this work reports the first observation of a quantum effect termed as nonlocal modulation. We have experimentally shown how distant modulators, when correlated in the frequency domain,  may interfere constructively or destructively.  Though this work has dealt with the effects of synchronously-driven sinusoidal modulators, a more general statement for nonlocal modulation is that phase modulation in Channel 1 of the form $\exp[i \Phi_1(t)]$ acts cumulatively with modulation $\exp[i \Phi_2(t)]$ in Channel 2 so as to produce a frequency domain correlation proportional to the square of the Fourier transform of  $\exp \{ i [\Phi_1(t)+\Phi_2(t)] \}$. For this relation to hold, it is required that $\Phi_1(t)$ and $\Phi_2(t)$ both vary slowly as compared to the temporal width of the biphoton wave function. 

\begin{acknowledgments}
The authors thank Irfan Ali-Khan for helpful discussions.  This work was supported by the U.S. Air Force Office of Scientific Research, the U.S. Army Research Office, and the Defense Advanced Research Projects Agency.
\end{acknowledgments}

\bibliographystyle{apsrev}

\end{document}